\documentclass{optica-article}

\journal{optcon}

\articletype{Research Article}
\usepackage{physics}
\usepackage[caption=false]{subfig}
\usepackage{lineno}

\begin{document}

\title{Quantum signatures in a quadratic optomechanical heat engine with an atom in a tapered trap}

\author{Mohsen Izadyari,\authormark{1,*} Mehmet Öncü,\authormark{2} Kadir Durak,\authormark{2} and Özgür E. Müstecaplıoğlu\authormark{1,3}}

\address{\authormark{1}Department of Physics, Ko\c{c} University, Istanbul, 34450, Turkey\\
\authormark{2}Department of Physics, \"{O}zye\u{g}in University,  Istanbul, 34794, Turkey\\
\authormark{3}T\"{U}B\.{I}TAK Research Institute for Fundamental Sciences, Gebze, 41470, Turkey}

\email{\authormark{*}mizadyari18@ku.edu.tr} 



\begin{abstract} We investigate how quantum signatures can emerge in a single atom
heat engine consisting of an atom confined in a tapered trap and
subject to hot and cold thermal reservoirs. A similar system was
realized experimentally in Ref.~\cite{singleAtom}. We model such a system using a quadratic optomechanical model and identify an effective Otto cycle in the system's dynamics. We compare the engine's performance in the quantum and classical regimes by evaluating the power dissipated. We find that lowering the temperature is insufficient to make the single atom engine of Ref.~\cite{singleAtom} a genuine quantum-enhanced heat engine. We show that it is necessary to make the trap more asymmetric and confined to ensure that quantum correlations cause an enhancement in the power output.
\end{abstract}

\section{Introduction}
Quantum heat engines (QHEs) are machines harnessing thermal energy using quantum working substances~\cite{QHEScovil,QHELevy}. While the three-level maser, a quantum device that amplifies microwave radiation, was recognized as a QHE several decades ago~\cite{QHE3masers}, systematic studies of QHE have attracted much attention quite recently ~\cite{decQHE01,decQHE02,decQHE03,decQHE04,decQHE05,decQHE06,decQHE07,decQHE08,decQHE09,decQHE10,decQHE11,decQHE12,etc01,etc02,etc03,physicaHE05,gershon02,etc04,etc05,etc06,etc07,physicaHE03,physicaHE04,LasercoolingBook,singleAtom02,flyeheel01,flyeheel02,Jiteng,Gershon01,Gershon03,physicaHE01,physicaHE02}. A pioneering experiment was the demonstration of a piston-type (Otto) engine using a single trapped ion~\cite{singleAtom}. Due to the relatively small energy gaps of the trapped ion at the involved temperatures, genuine, profound quantum effects in the single-atom piston were assumed to be non-existent. Here, we ask if it is sufficient to lower the temperature to bring the experimental setup to the quantum regime or if it is necessary to make more modifications. To answer this question, we use a higher-order, so-called quadratic, optomechanical model of the interaction between the motional degrees of freedom of the atom arising due to the tapered trap geometry of the experiment. 

It was intuitively expected that an optomechanical-type mutual interaction effectively emerges when an atom is confined in a funnel-shaped trap~\cite{singleAtom}. Here, we explicitly verify this intuition, but surprisingly, we find an additional degenerate three-wave mixing type coupling. It is a well-known interaction for generating squeezed states~\cite{newIntro01,newIntro02,newIntro03}. After that, we explore the signatures of the quantum squeezing in the system by differentiating classical and quantum correlations via solving master and classical stochastic Langevin equations separately. In a single atom system trapped in a tapered trap, we find that quantum correlations arise by increasing coupling strength between axial and radial modes. These quantum correlations can significantly enhance the dissipated power relative to the classical correlations. According to our model, the coupling strength depends on the trap's geometry and frequencies asymmetry. So, we indicate that the dissipated power can be enhanced by increasing the trap asymmetry and reducing the trap radius. The results show that lowering the temperature of the experiment~\cite{singleAtom} per se is not sufficient to bring the single-ion engine to the quantum regime. As a result, making the trap geometry more asymmetric and the atom more confined at low temperatures make quantum squeezing type correlations more significant for a remarkable performance increase than the classical stochastic engine. Accordingly, we conclude that, in addition to lowering the temperature, the original single-atom heat engine setup in Ref.~\cite{singleAtom} can be converted into a genuine quantum-enhanced heat engine in a more asymmetric and confined configuration. 

This paper is organized as follows. In Sec.~\ref{sec:model}, we describe our Hamiltonian model to describe the experimental ion in a tapered trap system in Ref.~\cite{singleAtom}. In Sec.~\ref{sec:results}, we investigate the dynamics of the system using the master equation approach. We identify an effective Otto cycle for the radial mode in Sec.~\ref{sec:QHE}. Sec. \ref{sec: classical} compares the stochastic and quantum dissipated power outputs through the axial mode. We conclude in Sec.~\ref{sec:conclusion}.
\section{\label{sec:model} Model System}

\begin{figure}[t!]
	\centering
	\includegraphics[width=4.2cm]{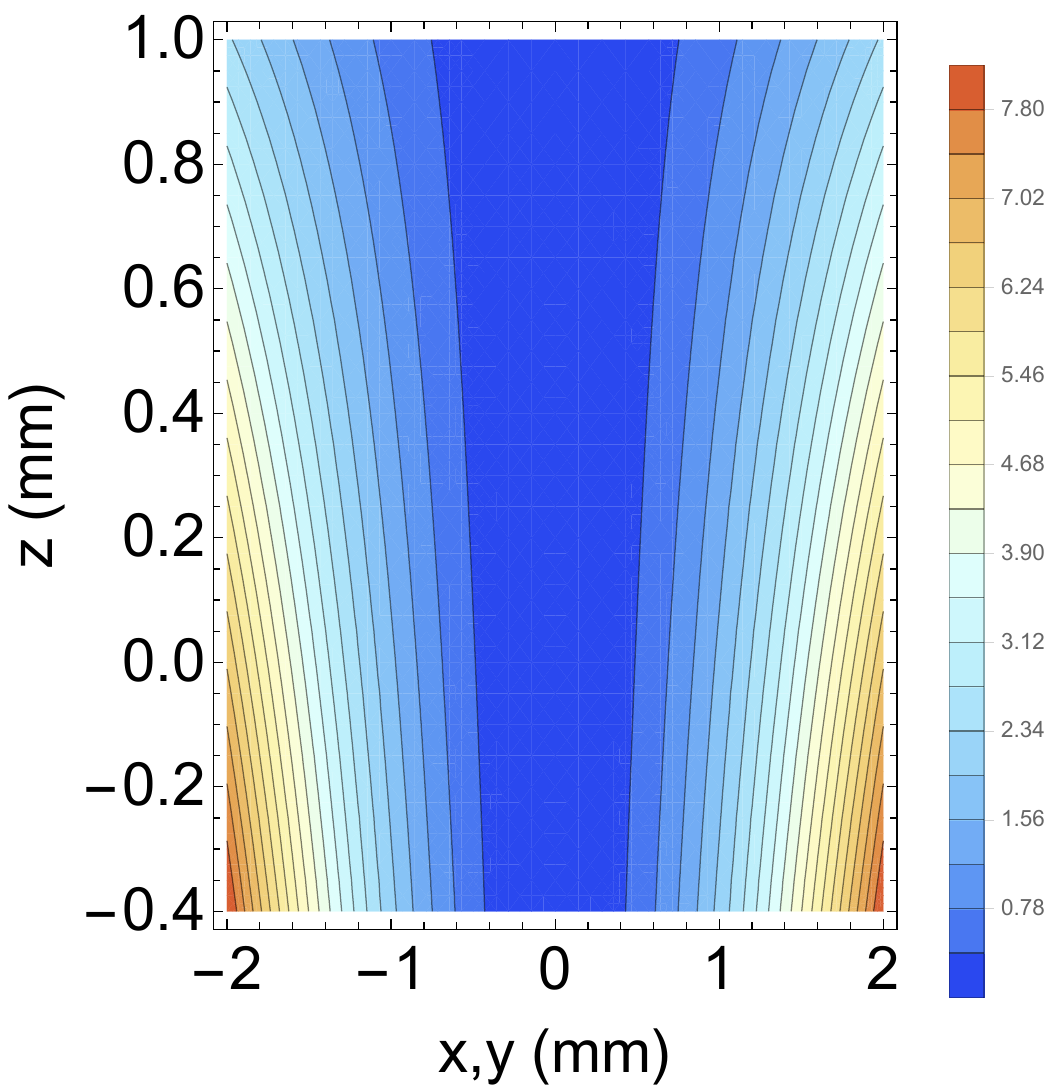} \hfil
	\includegraphics[width=4.2cm]{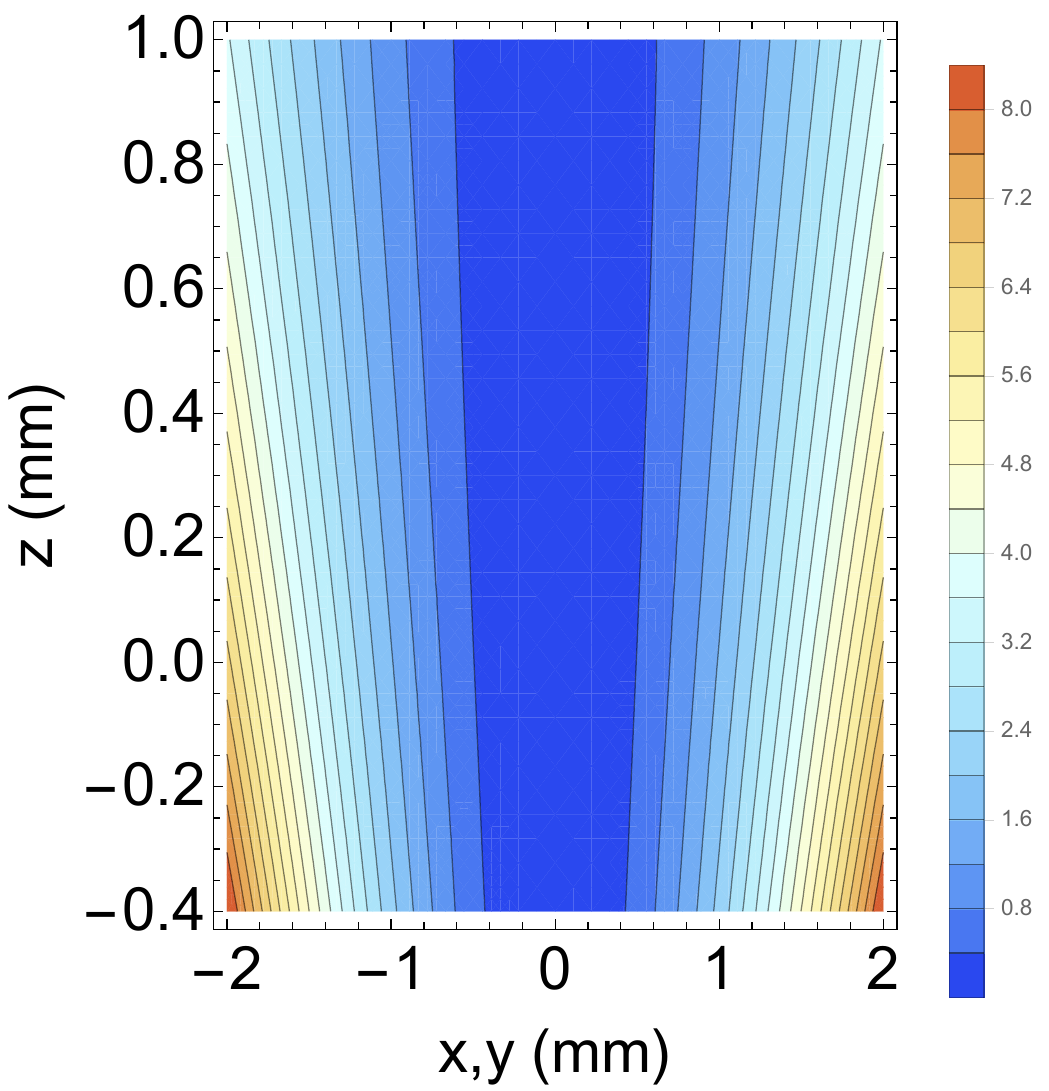}\par\medskip
	\includegraphics[width=4.2cm]{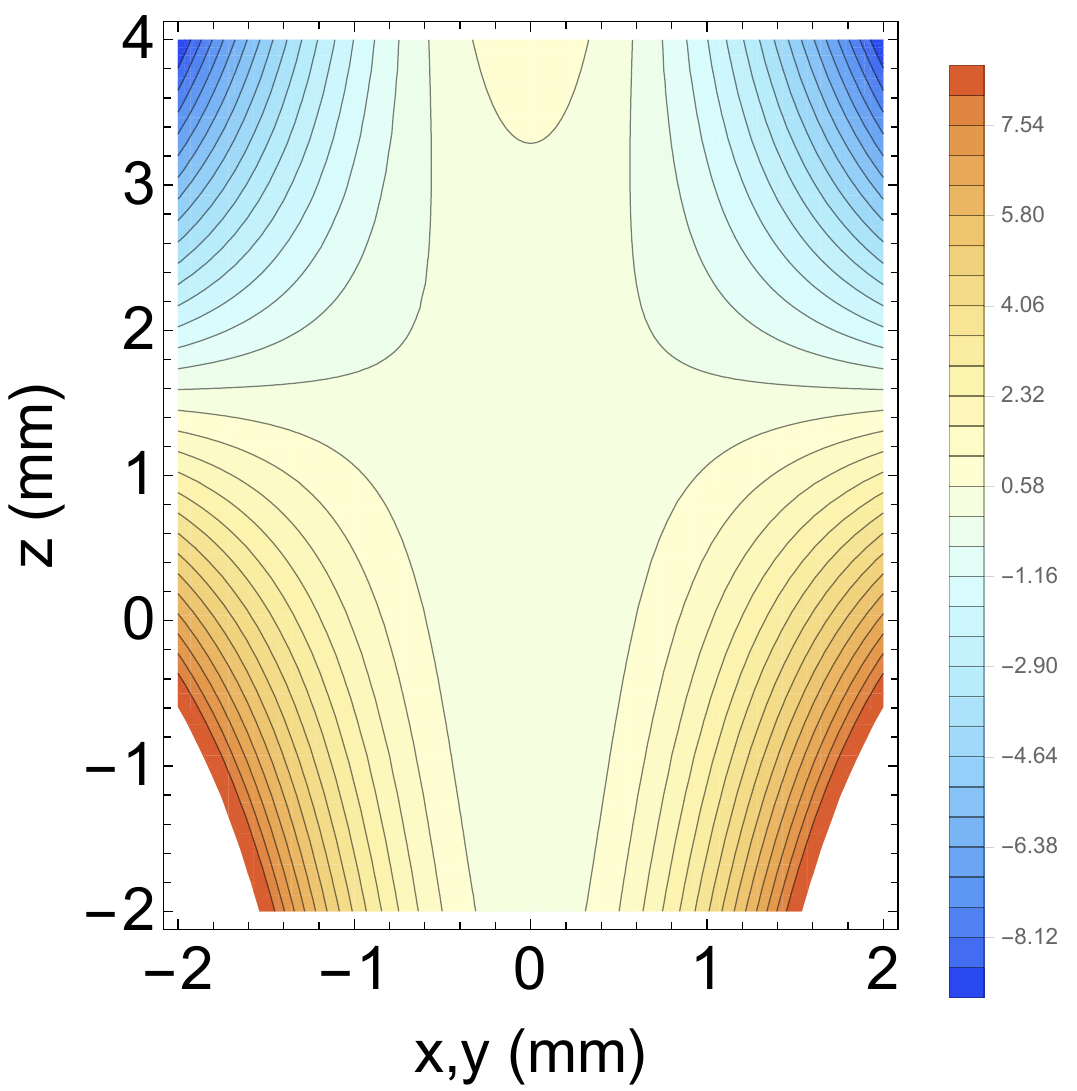} \hfil
	\includegraphics[width=4.2cm]{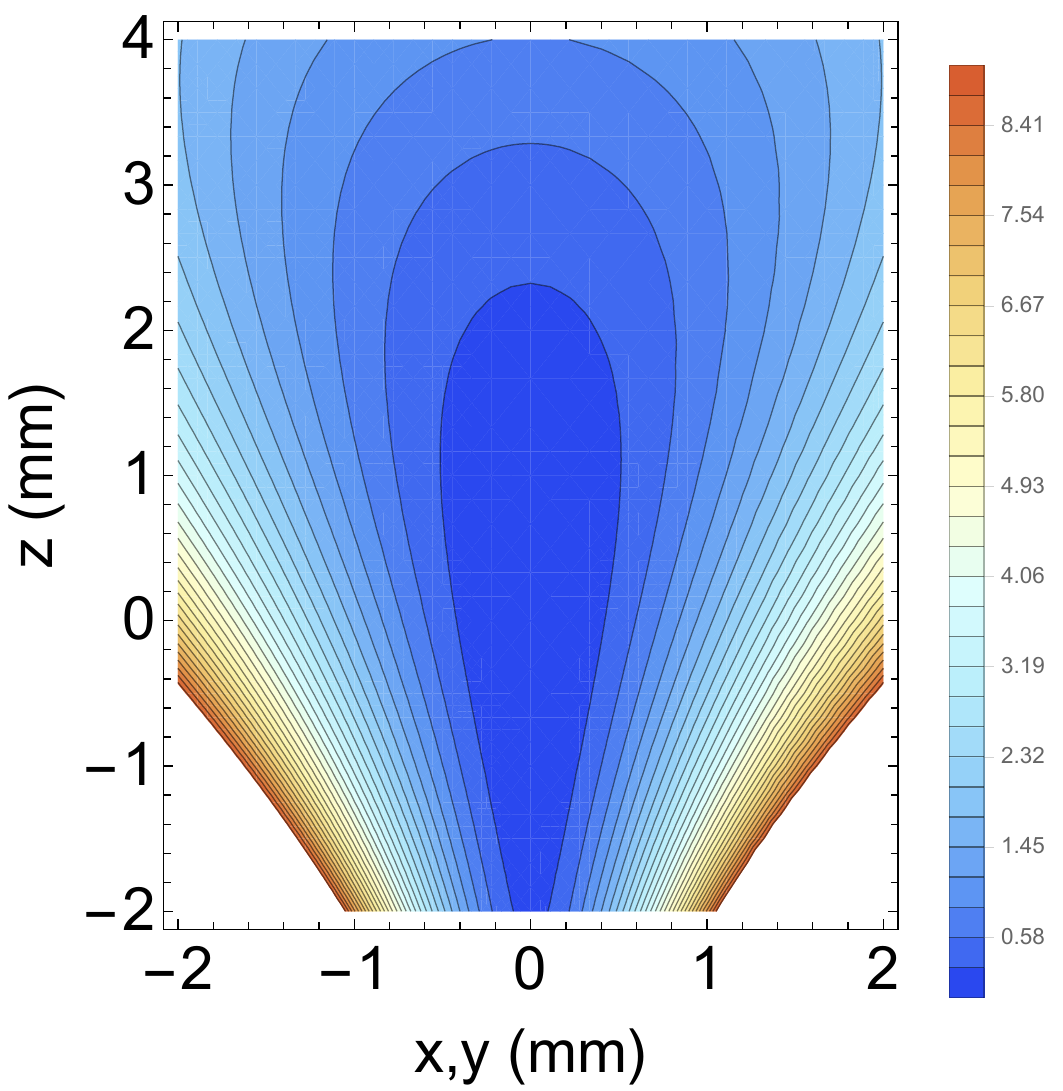}
	\caption{Cross sections of the approximate (Left panels)
		and exact (Right panels) funnel-shaped potential (in units
		of eV) of the tapered trap geometry. The approximation requires
		$z \tan \theta \ll r_0$ where $\theta = 30^o$ is the tapered trap angle and $r_0 = 1.1$ mm is the trap radius at $z = 0$. The calculations are done for the
		Calcium ion of mass $m = 40$u in atomic unit $u = 1.66\times10^{-27}$ kg.
	}
	\label{fig:trap pot}
\end{figure}

Single ion-heat engine realized in Ref.~\cite{singleAtom} consists of a Calcium ion of mass $m$ in a tapered trap geometry. In the axial ($z$) direction the trap frequency is $\omega_z$. In the directions, $x,y$, the trap frequencies are given by
\begin{equation} \label{eq: radial-freq}
    \omega_{\alpha} = \frac{\omega_{0\alpha}}{(1+\epsilon z)^2},
\end{equation}
where $\epsilon := (\tan \theta)/r_0$. Here, $r_0 = 1.1$ mm is the radius of the trap at $z =0$ and $\theta = 30^o$ is the trap angle. Trap frequencies along x and y axes are given as $\omega_{0x}$ and $\omega_{0y}$, respectively. It is found that
during the heat engine operation, the ion makes axial coherent oscillations with amplitude in the order of micrometers \cite{singleAtom}.
Accordingly, we can consider $\epsilon z \ll 1$ for the description of
heat engine dynamics. In this limit, the trap potential,
\begin{equation}\label{eq:potential001}
    U(x,y,z) = \frac{m}{2} (\omega_x^2 x^2 + \omega_y^2 y^2 + \omega_z^2 z^2),
\end{equation}
can be approximated by
\begin{equation}\label{eq:potential01}
V(r,y)=\frac{m}{2}\omega_{r}^2 r^2 (1-g z) +\frac{m}{2}\omega_{z}^2 z^2,
\end{equation}
where $r^2 = x^2 + y^2$, and $g = 4\epsilon \approx 2.1$ (1/mm). We
neglected the small difference between $\omega_x$ and $\omega_y$ and replace
them with the mean radial trap frequency, $\omega_r := (\omega_{0x} + \omega_{0y})/2$ \cite{singleAtom}. The trap potential $U(x, y, z)$ (using Eq.~(\ref{eq: radial-freq}) to define $\omega_x$ and $\omega_y$) and its approximation
by $V (r, z)$ are compared in Fig.~\ref{fig:trap pot} (right  and left panels respectively). 
According to Fig.~\ref{fig:trap pot} (upper panels), the exact potential can be superseded by approximated potential for small values of z.
\par
\begin{figure}[t!]
	\centering
	\includegraphics[width=8.6cm]
	{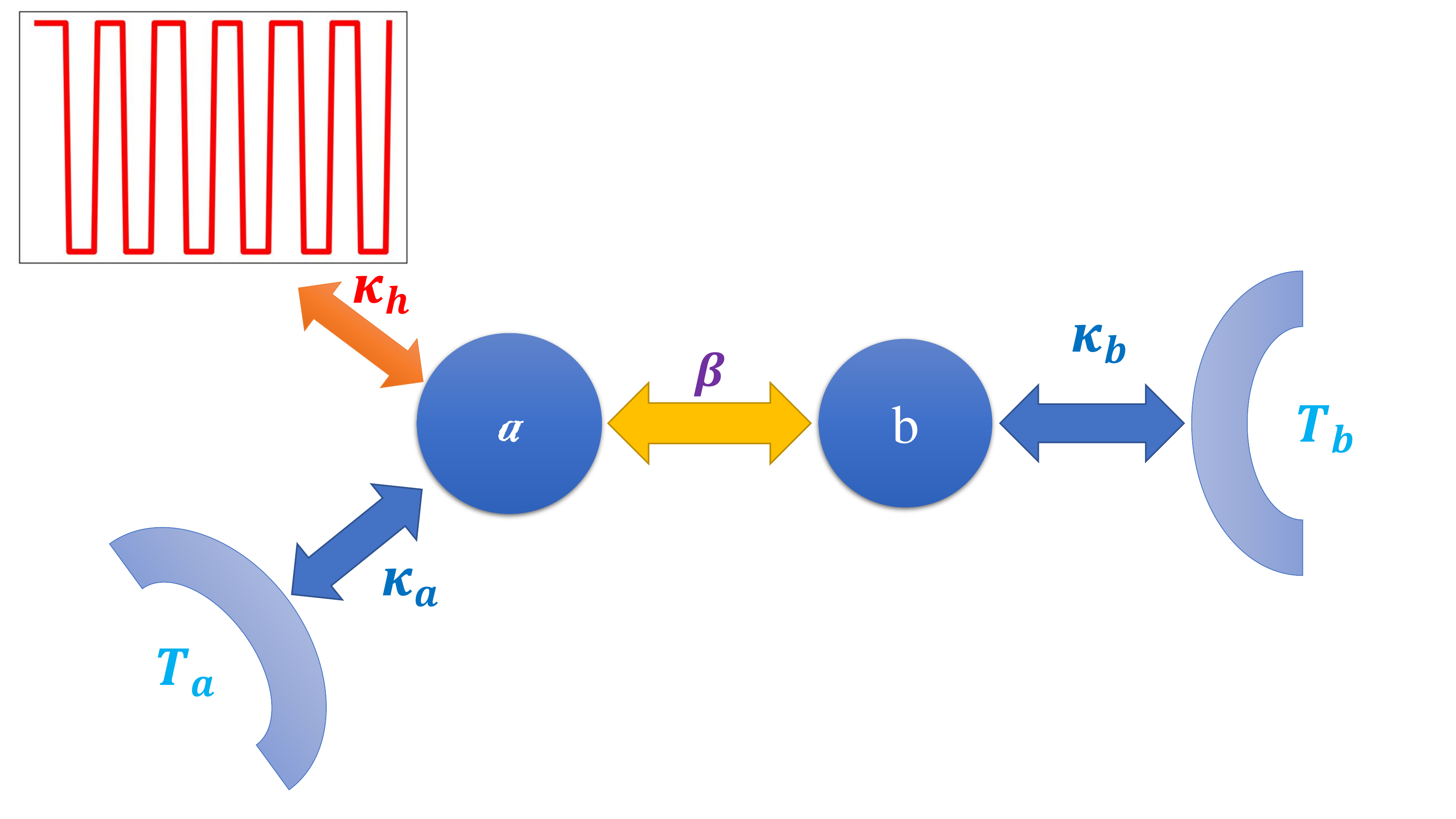}
	\caption{\label{fig:system}Diagram of the heat engine system based on the radial (a) and axial (b) vibrational modes whose coupling is characterized by the parameter $\beta$. Radial mode is coupled to two thermal baths at temperatures $T_h$ and $T_a$, where the coupling to the bath at temperature $T_h$ is periodically turned on and off, and the axial mode is coupled to a bath at temperature $T_b$. The background environment common to both modes is not shown.}
\end{figure}

 The trapped ion is always subject to a cooling laser yielding the effect of a cold thermal bath. An electrical white noise is applied to the radial degree of freedom for periodical heating as illustrated in Fig.~\ref{fig:system}. The heat engine system with two radial ($a$) and axial ($b$) vibrational modes (coupled by $\beta$) is shown in Fig.~\ref{fig:system}. Accordingly, two thermal baths at temperatures $T_h$ and $T_a$ couple to radial mode, and the axial mode is coupled to a bath at temperature $T_b$. The coupling between system and thermal bath at temperature $T_h$ is periodically turned on and off. The engine cycle is examined classically in Ref.~\cite{singleAtom}, which is justified for $k_B T \gg \hbar \omega_r$, where $T$ is considered as the working temperatures of the engine.
Indeed, for all temperatures of interest, including the high-temperature condition $T_c \gg \hbar \omega_r$ is satisfied.
On the other hand, the particular coupling of the axial and radial degrees of freedom in Eq. (\ref{eq:potential01}) contains a quadratic nonlinearity in $r$. From the quantum mechanical point of view, such a nonlinearity could yield a quantum noise squeezing effect. 
\par Let us express the quantum model of the center-of-mass (CM) motion, in the form
\begin{align} \label{eq:eq01}
H_0 & = \frac{P_r^2 + P_z^2}{2M} + V(r,z).
\end{align}
Here, $P_r$ and $P_z$ stand for the components of the CM momentum operator, with $P_r^2 =  P_x^2 + P_y^2$. We apply the harmonic-oscillator quantization $r =
(\hbar/m \omega_r)^{1/2} \hat q_r$, $z = (\hbar/m \omega_z)^{1/2} \hat q_z$, $P_r = (\hbar m \omega_r)^{1/2} \hat p_r$,
and $ P_z = (\hbar m \omega_z)^{1/2} \hat p_z$ with 
$\hat p_r = i(\hat a^{\dagger}-\hat a)/\sqrt{2}$, 
$\hat p_z = i(\hat b^{\dagger}-\hat b)/\sqrt{2}$,
$\hat q_r = (\hat a^{\dagger} +\hat a)/\sqrt{2}$, and $\hat q_z = (\hat b^{\dagger} +\hat b)/\sqrt{2}$ to write the Hamiltonian~(\ref{eq:eq01})
in terms of the vibrational phonon operators, (we take $\hbar = 1$)
\begin{align} \label{eq:ham_CM}
\hat H_\text{cm} & = \omega_r \hat n_r + \omega_z \hat n_z - \frac{\beta}{4} (\hat a^2 + \hat a^{\dagger 2} + 2 \hat a^{\dagger}\hat a + 1)(\hat b + \hat b^{\dagger}),
\end{align}
where we have dropped the constants $\omega_r /2$ and $\omega_z /2$. $\beta \equiv g \omega_r \sqrt{\hbar/2m \omega_z}$ is the coupling coefficient of $r$ and $z$ degrees of freedom; $\hat n_r = \hat a^{\dagger} \hat a$, and $\hat n_z = \hat b^{\dagger} \hat b$ are the phonon number operators in $r$ and $z$ directions, respectively. We remark that the coupling coefficient $\beta$ depends on trap geometry through the coupling constant g.

\par The Hamiltonian~(\ref{eq:ham_CM}) can be compared to an optomechanical-like Hamiltonian, but in addition to the occupation of mode $a$, quantum squeezing-type correlations can induce additional pressure on mode-$b$ to produce coherent displacement or useful work. To isolate the effects of different coupling mechanisms in Eq.~(\ref{eq:ham_CM}), we simulate each of them using the separate models,
\begin{align} 
H_{\text{om}} & = \omega_r \hat n_r + \omega_z \hat n_z - \frac{\beta}{2}  \hat a^{\dagger}\hat a (\hat b + \hat b^{\dagger}), \label{eq:Resonator H01}\\ 
H_{\text{class}} & = \omega_r \hat n_r + \omega_z \hat n_z - \frac{\beta}{4} (\hat b + \hat b^{\dagger}), \label{eq:Resonator H02} \\ 
H_{\text{sq}} & = \omega_r \hat n_r + \omega_z \hat n_z - \frac{\beta}{4} (\hat a^2 + \hat a^{\dagger 2})(\hat b+\hat b^{\dagger}). \label{eq:Resonator H03}
\end{align}
Eq.~(\ref{eq:Resonator H01}) describes a pure optomechanical coupling where only the occupation of mode-$a$ produces pressure on mode-$b$. Modes are uncoupled in Hamiltonian~(\ref{eq:Resonator H02}), but there is still an effective classical drive on mode-$b$. Eq.~(\ref{eq:Resonator H03}) injects coherence in mode-$b$ depending on the squeezing of mode $a$.
\par In the next section, we will simulate the system based on
total Hamiltonian Eq.~(\ref{eq:ham_CM}) as system $A_0$, and each of the models in Eqs. (\ref{eq:Resonator H01}-\ref{eq:Resonator H03}) as systems $A_1$, $A_2$ and $A_3$, respectively.
Let’s emphasize that optomechanical coupling can not emerge per se but is always accompanied by the other classical and squeezed drive terms, and each coupling contributes with comparable or the same strengths. Our separation is artificial and only for the purpose of shedding light on the individual effects of each term on the coherent dynamics of mode-b, in particular on the work output of the single ion heat engine. We note that Ref. \cite{etc06} assumes Gaussian thermal distribution to model
the spatial distribution of mode-a. This description would be closer to the classical treatment of our optomechanical model of $A_1$ given by Eq.~(\ref{eq:Resonator H01}). 

\section{Results}\label{sec:results}
In this section, we introduce the parameters we used in our simulations and describe the open quantum system dynamics of the system in terms of a master equation.
\subsection{Parameters}\label{sec:parameters}

The experiment~\cite{singleAtom} has been performed at high temperature (in the order of mK). As the usual route to the quantum regime is to reduce the temperature, we consider the temperature in the order of the $\mu$K. However, we find that lowering the temperature is insufficient to see the system's quantum effects. It is required to demonstrate the system can harness more power than its stochastic counterpart to be classified as a genuine quantum (enhanced) heat engine~\cite{quantumSignature}. In addition to reducing the temperature, increasing $\beta$ plays a crucial role in achieving clear signatures of quantum enhancement in the heat engine operation with a single ion in a tapered trap. It is necessary to exploit quantum squeezing type correlations to enhance power output relative to the case where there are only classical correlations.

In the following, we take $T_h = 166~\mu$K and $T_a = T_b = 4~\mu$K while we suppose the atom's initial temperature is $T_0 = 10~\mu$K. Also, the radial and axial frequencies are taken $\omega_r/ 2\pi = 1$ MHz and $\omega_z /2\pi = 50$ kHz, respectively. According to the definition of the coupling coefficient ($\beta \propto \tan \theta / r_0$), we consider high trap asymmetry by taking $\theta = \pi/4$ and considering more confined traps with $r_0 \lesssim 20~\mu$m~\cite{microtrap01,microtrap02} to make the coupling stronger. 
For example, for $r_0 \approx 2~\mu$m  and using the aforementioned parameters, we find $\beta/2\pi = 100$ kHz. We emphasize that these parameters are not arbitrary but
taken as an example from a range of values for which the temperatures are sufficiently low and squeezing type coupling is sufficiently large to get into the quantum enhanced
regime of engine operation.

The dissipation rates of mode $a$ to the cold and hot baths are considered to be $\kappa_a /2\pi = \kappa_h /2\pi = 200$ kHz, where $\kappa_h$ is taken as a time-dependent square wave $\kappa_h (t):= \kappa_h u(t)$. The time-dependent part is $u(t)=1$ for the heating and otherwise $u(t)=0$. Also, the dissipation rate of mode $b$ to the cold bath is $\kappa_b/2\pi = 50$ kHz. 
\subsection{The dynamics of the system}\label{sec:dynamics}

According to Eq.~(\ref{eq:ham_CM}), the model effectively describes two nonlinearly coupled single-mode resonators. 
We assume a microwave white noise and a cooling laser are applied to the mode $a$, as shown in the experiment of Ref.~\cite{singleAtom}. This continuous electrical noise and cooling laser plays the roles of two cold and heat baths at temperatures $T_a$ and $T_h$ such that $T_h > T_a$. An extra cooling laser beam is introduced to mode $b$ to keep the axial mode's oscillations in control by tuning its parameters on
purpose to get a limit cycle at temperature $T_b$. The background environment with temperature $T_0$ has been ignored during the dynamics by assuming the coupling coefficient of background environment with mode $b$ and $a$ less than $\kappa_j$ where $j \in \{h, a, b\}$ \cite{SemiQHE}. 
The dynamics of the density matrix $\rho$ of the two-resonator system is determined by the master
equation~\cite{QopenSystemBook,QOscully,localMasterEq}
\begin{align} \label{eq:master eq 01}
\frac{d\rho}{dt} & = -i[\hat H_{\lambda}, \rho] \nonumber\\ &+ \kappa_a (\Bar{n}_a + 1) D[\hat a] + \kappa_a \Bar{n}_a D[\hat a^{\dagger}] \nonumber \\
& + \kappa_h(t) (\Bar{n}_h + 1) D[\hat a] + \kappa_h(t) \Bar{n}_h D[\hat a^{\dagger}] \nonumber \\ &+ \kappa_b (\Bar{n}_b + 1) D[\hat b] + \kappa_b \Bar{n}_b D[\hat b^{\dagger}],
\end{align}
where $\lambda =\text{cm, om, class, sq}$ represents different models and $D[\hat \alpha]=(1/2)(2\hat \alpha \rho \hat \alpha ^{\dagger} - \hat \alpha^{\dagger} \rho \hat \alpha - \rho \hat \alpha ^{\dagger} \hat \alpha)$ refers to the Lindblad dissipator superoperators with $\hat \alpha = \hat a, \hat b$. 
Mean number of phonons in mode $a$ and $b$ at the bath temperature $T_u$ and $T_b$ is denoted by $\Bar{n}_u$ with $u \in \{a,h\}$ and $\Bar{n}_b$, given by the Planck distribution (we take $k_B = 1$)
\begin{align} \label{eq:planck dis}
    \Bar{n}_u &= \frac{1}{\exp(\omega_r / T_u) - 1}, \nonumber \\ 
    \Bar{n}_b &= \frac{1}{\exp(\omega_z / T_b) - 1}.
\end{align}

We assume that before the cooling and heating lasers are applied, the ion and its vibrational modes are in thermal equilibrium with the background thermal environment at temperature
$T_0$. Initial state then can be expressed as a canonical thermal (Gibbs) state such that 
\begin{eqnarray} \label{eq:initState}
 \rho (0) & = \frac{1}{\text{Tr} [e^{-\hat H_{\lambda} /T_0}]}e^{-\hat H_{\lambda} /T_0},
\end{eqnarray}
and the initial states of the vibrational modes are determined by the reduced density matrices $\rho_b (0) = \Tr_b (\rho (0))$ and $\rho_a (0) = \Tr_a (\rho (0))$.
We solve the master equation to determine $\rho(t)$ for the given $\rho(0)$ using Python programming language and an open source quantum optics package QuTiP~\cite{Qutip}. Taking the partial trace of $\rho(t)$ over the degrees of freedom of mode-$a$, we find the reduced state of the mode-$b$ such that
$\rho_{b}(t)=\Tr_{a}[\rho(t)]$. Time dependence of the mean number of vibrational phonons in mode-$b$ is evaluated by 
$\langle \hat n_z \rangle = \Tr ( \rho_b(t) \hat b^{\dagger} \hat b)$. 
Similar calculations give dynamics of $\langle \hat n_r \rangle$, too. 
\begin{figure}[t!]
	\centering
	\includegraphics[width=8.6cm]
	{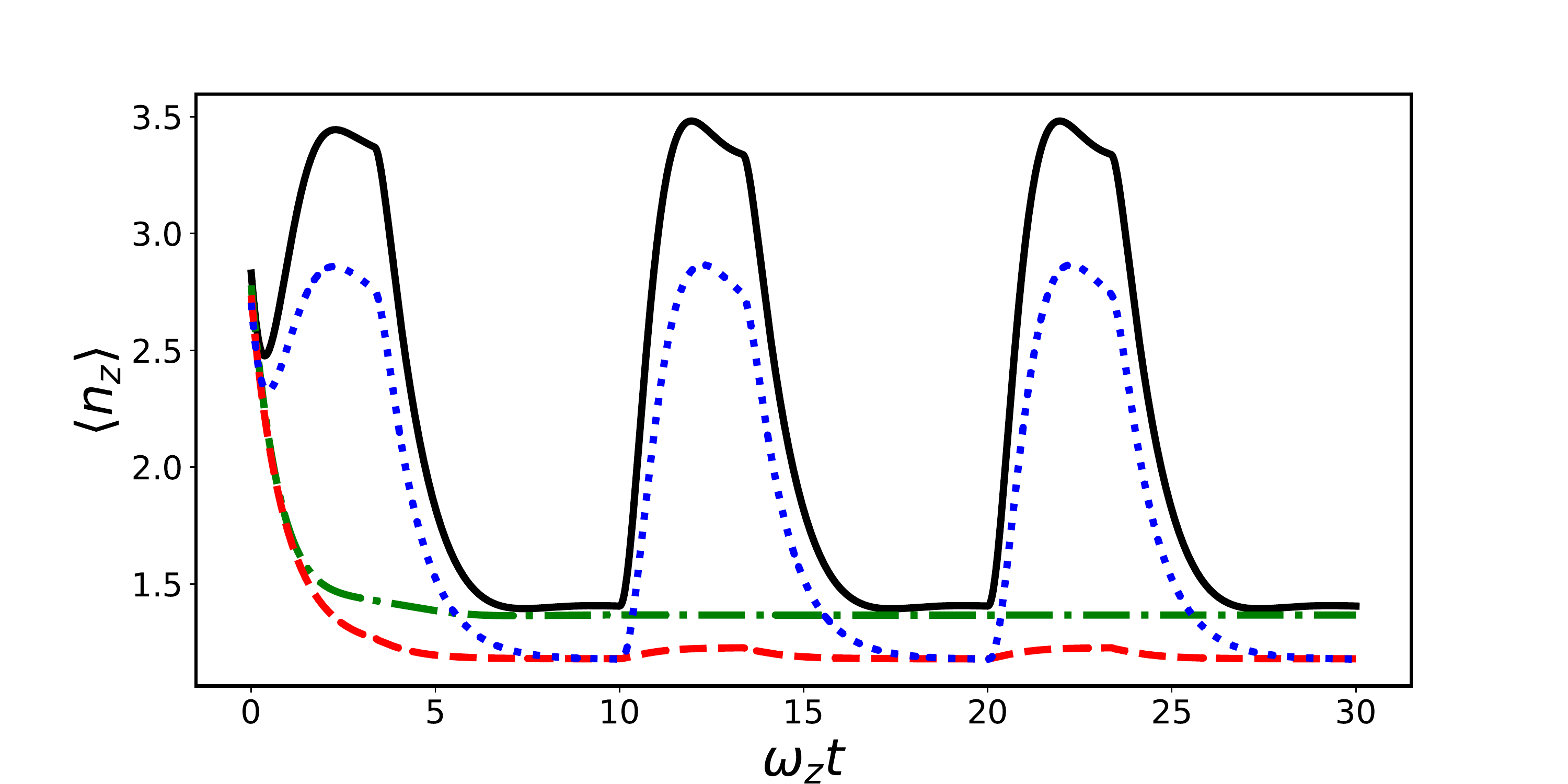}
	\caption{\label{fig:nb fig2}Dynamics of the mean number of
		vibrational phonons $\langle n_z \rangle$ in modes $b$ where time is scaled with $\omega_z$. The parameters used in the plots are taken to be $\omega_r/2\pi =1$ MHz, $\omega_z/2\pi = 50$ kHz, $\Bar{n}_h = 3$, and $\beta/2\pi = 100$ kHz. The solid black, dotted blue, dash-dotted green, and dashed red lines show the evolution of $\langle n_z \rangle$ under the models $A_0$, $A_1$, $A_2$, and $A_3$, respectively.
	}
\end{figure}

The evolutions of $\langle \hat n_z \rangle$ for the four models 
$A_0$ to $A_3$ are shown in Fig.~\ref{fig:nb fig2}. Since all the models $A_0$ to $A_3$ have different Hamiltonians, the initial conditions differ. Nevertheless, the differences in the initial value of $\langle \hat n_z \rangle$ are too small to be visible in Fig.~\ref{fig:nb fig2}.
Among all the terms in the overall interaction $A_0$, optomechanical type coupling $A_1$ is the only mechanism to increase the axial mode population. However, despite the qualitative agreement, $A_1$ per se cannot quantitatively explain the $\langle \hat n_z \rangle$ behavior. While the squeezing term
	$A_3$ cannot directly populate $\langle \hat n_z \rangle$; it plays a subtle role to
	increase $\langle \hat n_z \rangle$ beyond what is possible solely by $A_1$. 

\begin{figure}[t!]
	\centering
	\includegraphics[width=8.6cm]{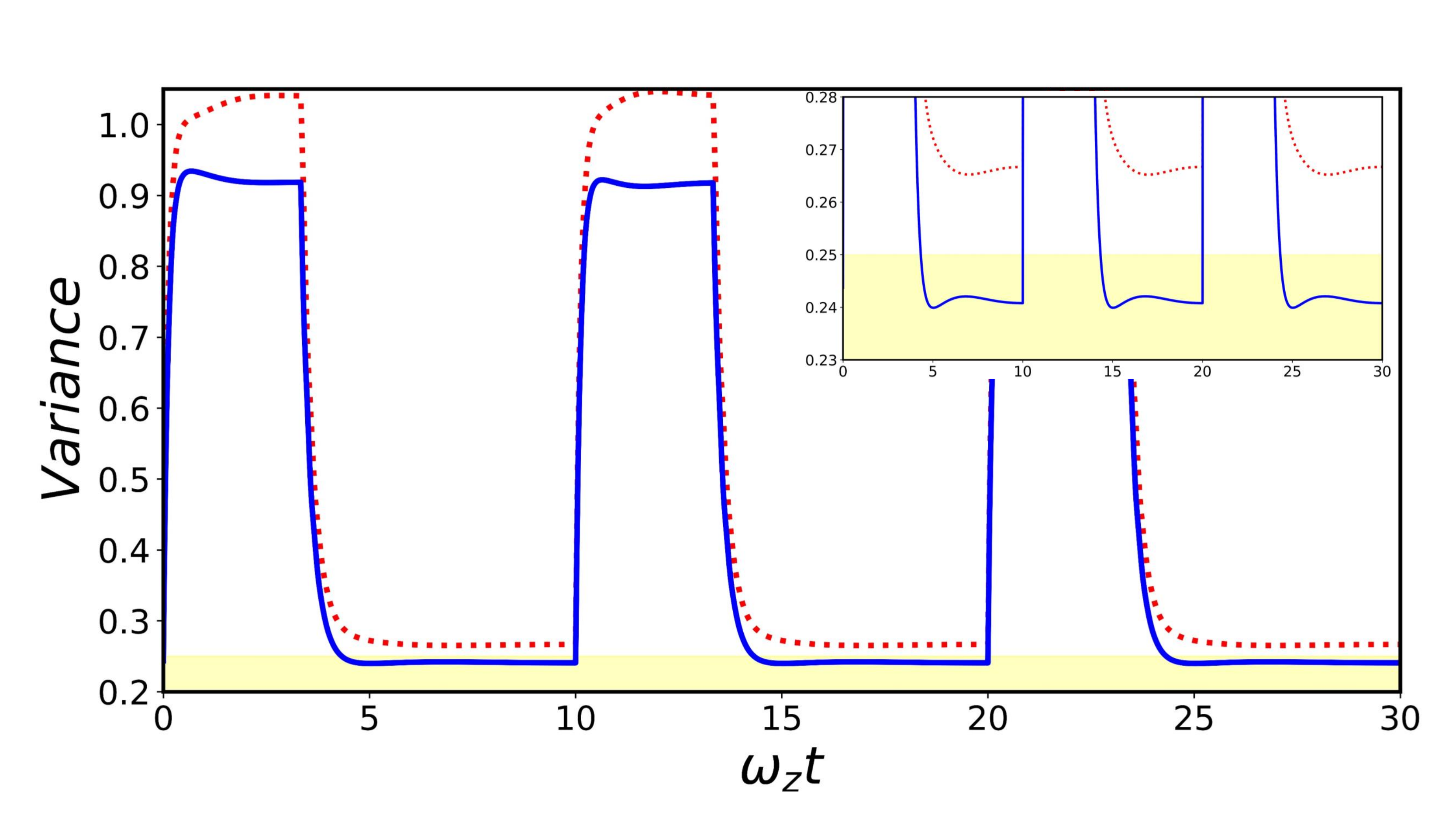}
	\caption{The time evolutions of the variances of the quadratures $(\langle \Delta \hat q_r \rangle)^2$ (red dotted line) and$(\langle \Delta \hat p_r \rangle)^2$ (blue solid line) of the radial mode. The inset magnifies the squeezing region where Variance $< 1/4$.}
	\label{fig:squeezing}
\end{figure}
Intuitively, the effect of squeezing-type terms $A_3$ in $A_0$ can be interpreted as an enhancement of vibrational radial pressure on the axial mode. 
When $A_3$ terms are present, the number of radial vibrational phonons during the heating stage is expected to increase so that the optomechanical-like coupling $A_1$ in $A_0$ can 
yield higher $\langle \hat n_z \rangle$. The more radial phonons
cause stronger pressure on the axial mode to transfer more energy or power. 
To appreciate how the quantum character
of the modes enters into this picture, let us look at the dynamics of $\langle \hat n_z \rangle$ according to the master equationn~(\ref{eq:master eq 01})
\begin{eqnarray}\label{eq:nz}
	\frac{d}{dt}\langle \hat n_z \rangle 
	=\frac{\beta}{\sqrt{2}}(\langle \hat q_r^2,\hat p_z\rangle+\langle \hat q_r^2\rangle\langle\hat p_z\rangle)
	-\kappa_b(\langle \hat n_z\rangle- \Bar{n}_b),
\end{eqnarray}
where $\langle \hat q_r^2,\hat p_z\rangle:=\langle \hat q_r^2,\hat p_z\rangle-\langle \hat q_r^2\rangle\langle\hat p_z\rangle$ is a quantum correlation function between the radial and axial modes. The dissipated power output of the engine is determined by $P_{\text{dis}} = \omega_z \kappa_b (\langle \hat{n}_z \rangle - \Bar{n}_b)$, which measures the net flux of energy dissipated into the environment~\cite{power}. In order to produce more power than a stochastic engine that can only have classical correlations, it is necessary for us to ensure that the
system has strong quantum correlations associated with the second moment of the displacement of the radial mode. Quantum character of the $\langle \hat q_r^2 \rangle$ can be examined in terms of the quadrature variances of the radial mode
\begin{eqnarray} \label{eq:dXdP}
(\langle \Delta \hat q_r \rangle)^2 &=&  Tr[\rho_a \hat q_r^2] - (Tr[\rho_a \hat q_r])^2, \\
(\langle \Delta \hat p_r \rangle)^2 &=& Tr[\rho_a \hat p_r^2] - (Tr[\rho_a \hat p_r])^2,
\end{eqnarray} 
where $\rho_a(t) = Tr_b[\rho(t)]$ is the reduced density
matrix of the radial mode. When a quadrature variance is less than its vacuum value of $1/4$, the state is said to be quadrature squeezed.

\begin{figure}[t]
    \centering
    \subfloat[]{{\includegraphics[width=0.28\linewidth]{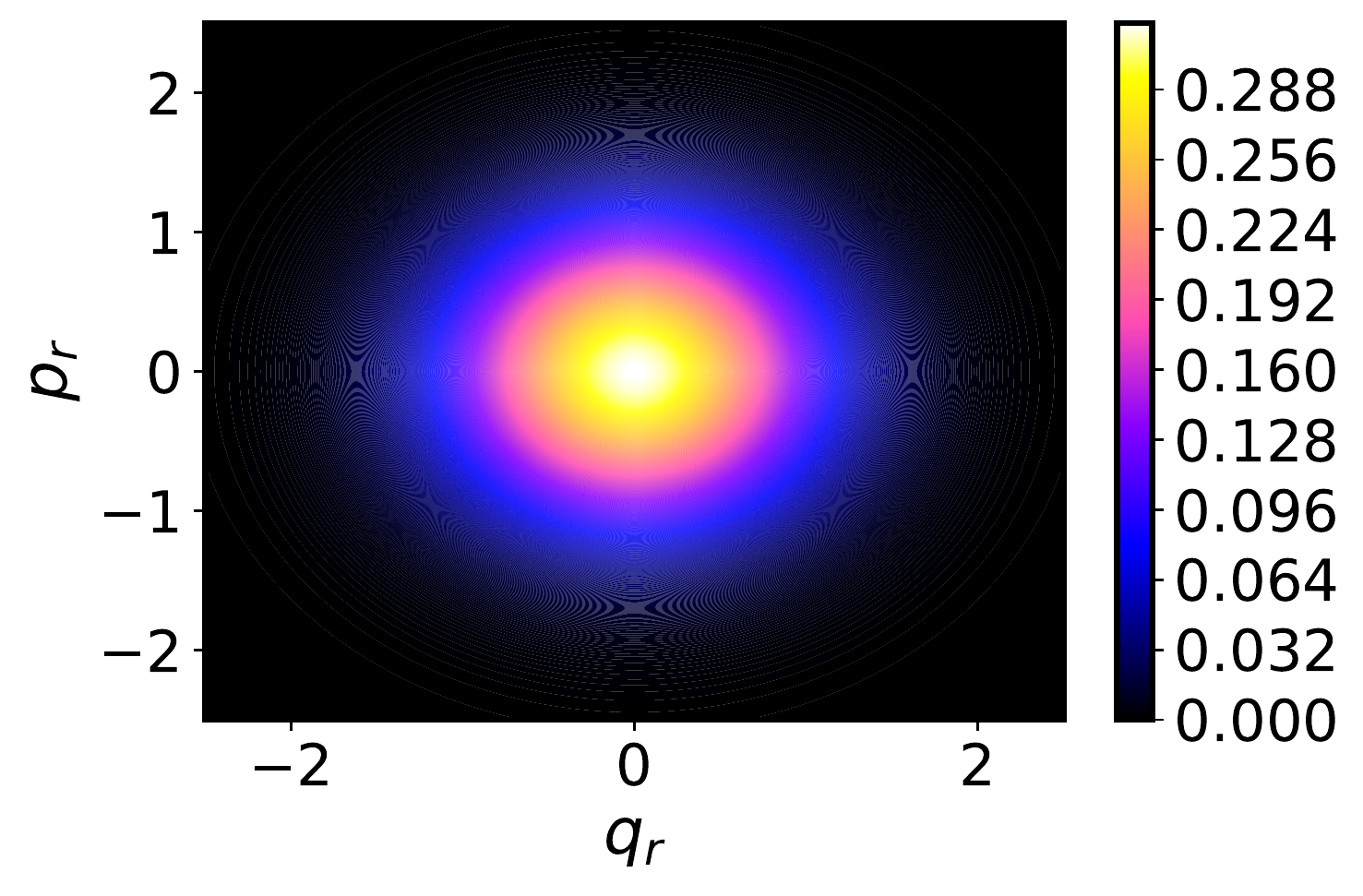} }}
    \qquad
    \subfloat[]{{\includegraphics[width=0.28\linewidth]{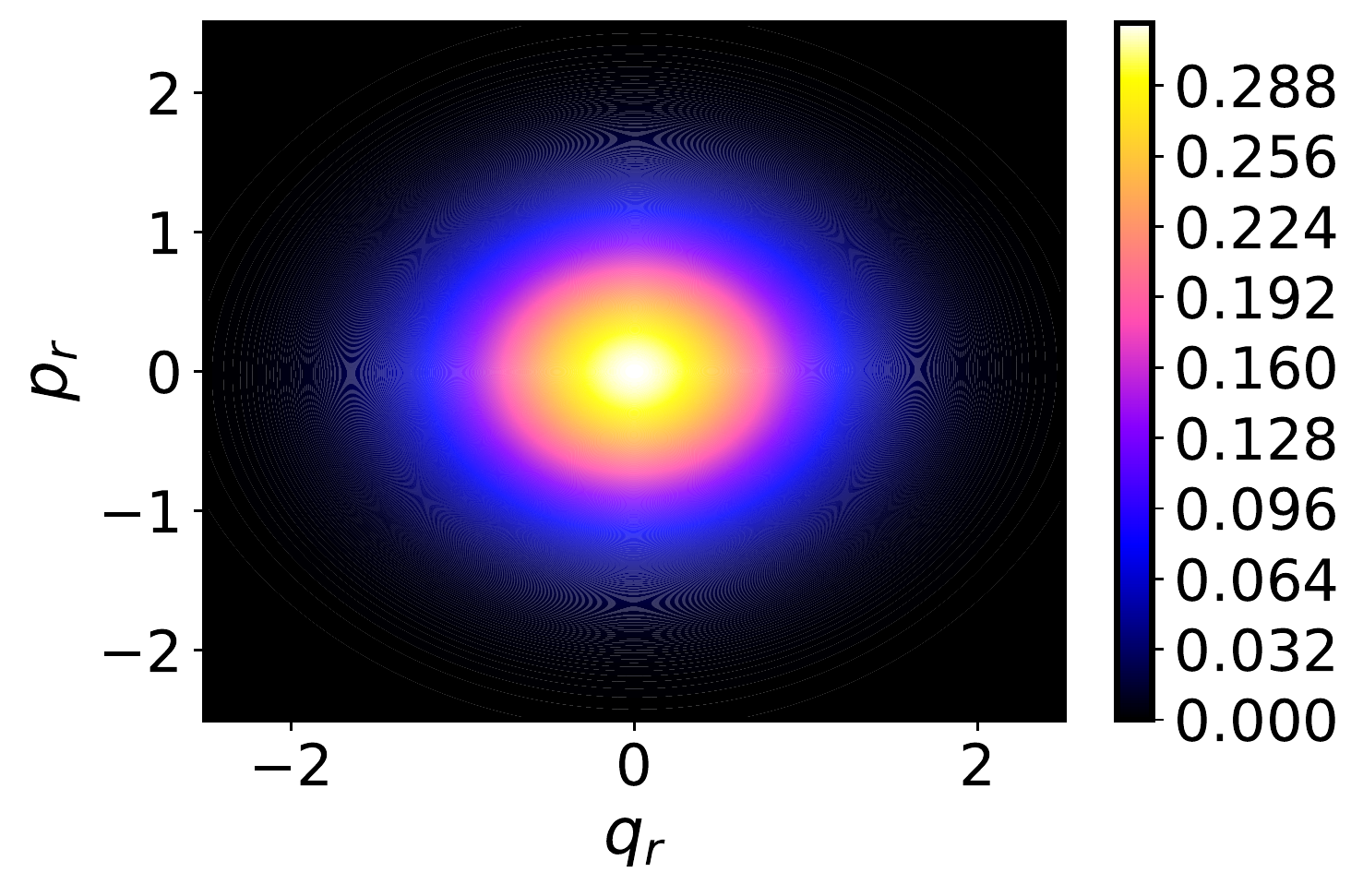} }}
    \qquad
    \subfloat[]{{\includegraphics[width=0.28\linewidth]{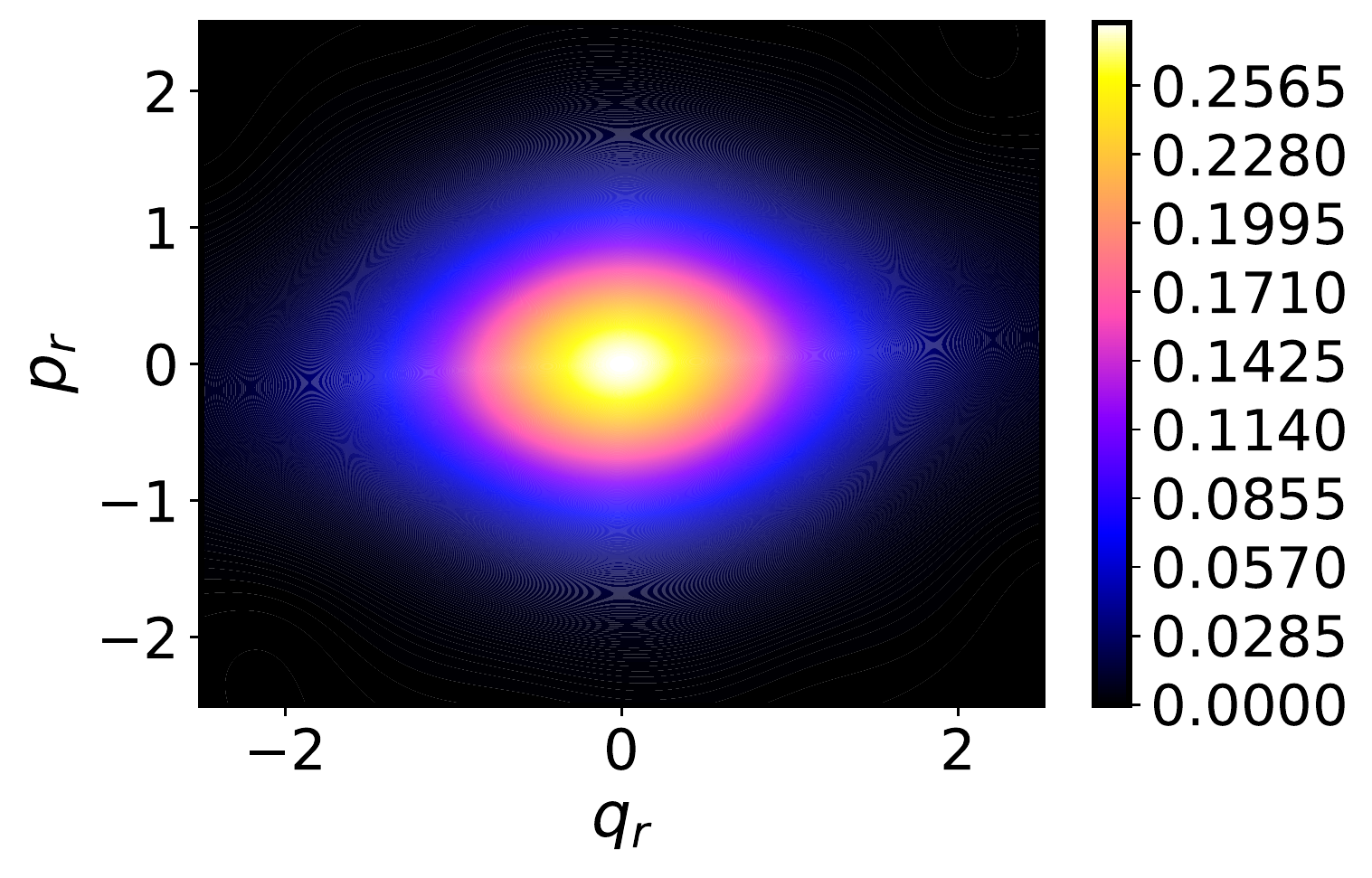} }}
    
    \caption{Colour online)(a)-(c) The Wigner functions in the $p_r$, $q_r$ field quadrature phase space of the radial mode plotted for (a) $\beta /2\pi = 10$ kHz, (b) $\beta /2\pi = 100$ kHz, (c) $\beta /2\pi = 200$ kHz. The other system parameters are $\omega_r/2\pi = 1$ MHz, $\omega_z/2\pi = 50$ kHz.}
    \label{fig:contour plot}
\end{figure}

The time evolutions of $(\langle \Delta \hat q_r \rangle)^2$ (red dotted line) and $(\langle \Delta \hat p_r \rangle)^2$ (blue solid line) are shown in Fig.~\ref{fig:squeezing}. 
The yellow shaded area, where Variance $< 1/4$, in Fig.~\ref{fig:squeezing} indicates that the radial mode starts each cycle in a squeezed thermal state.  
Fig.~\ref{fig:contour plot} plots the Wigner functions of the radial mode for different $\beta$, at the same scaled time after thermalization of the system, when the system dynamics is governed by the full model $A_0$. Wigner function of a thermal state in the quadrature phase space has circular contours, which are changed to ellipses	due to quadrature squeezing. 
In the Fig.~\ref{fig:contour plot}, we show that although the quadrature phase space of radial mode has circular contours for small $\beta$ (Fig.~\ref{fig:contour plot} (a)), it becomes to ellipses contours by enhancing $\beta$ (figs.~\ref{fig:contour plot} (b,c)).
So, for small $\beta$, the eccentricity is almost zero, and the squeezing is negligible even at low temperatures.
Accordingly, the dissipated power of the engine will be dominated and determined by the classical correlations only.
On the other hand, for large values of $\beta$, second order moments of the displacement operator should be treated quantum mechanically in Eq.~\ref{eq:nz} and hence to the calculation of the dissipated power. We present explicit calculation of
$P_{\text{dis}}$ for both quantum and classical stochastic treatments 
in Sec.~\ref{sec: classical}.


\section{Effective Otto Cycle For the Radial Mode}\label{sec:QHE}
\begin{figure}
    \centering
    \includegraphics[width=8.6cm]{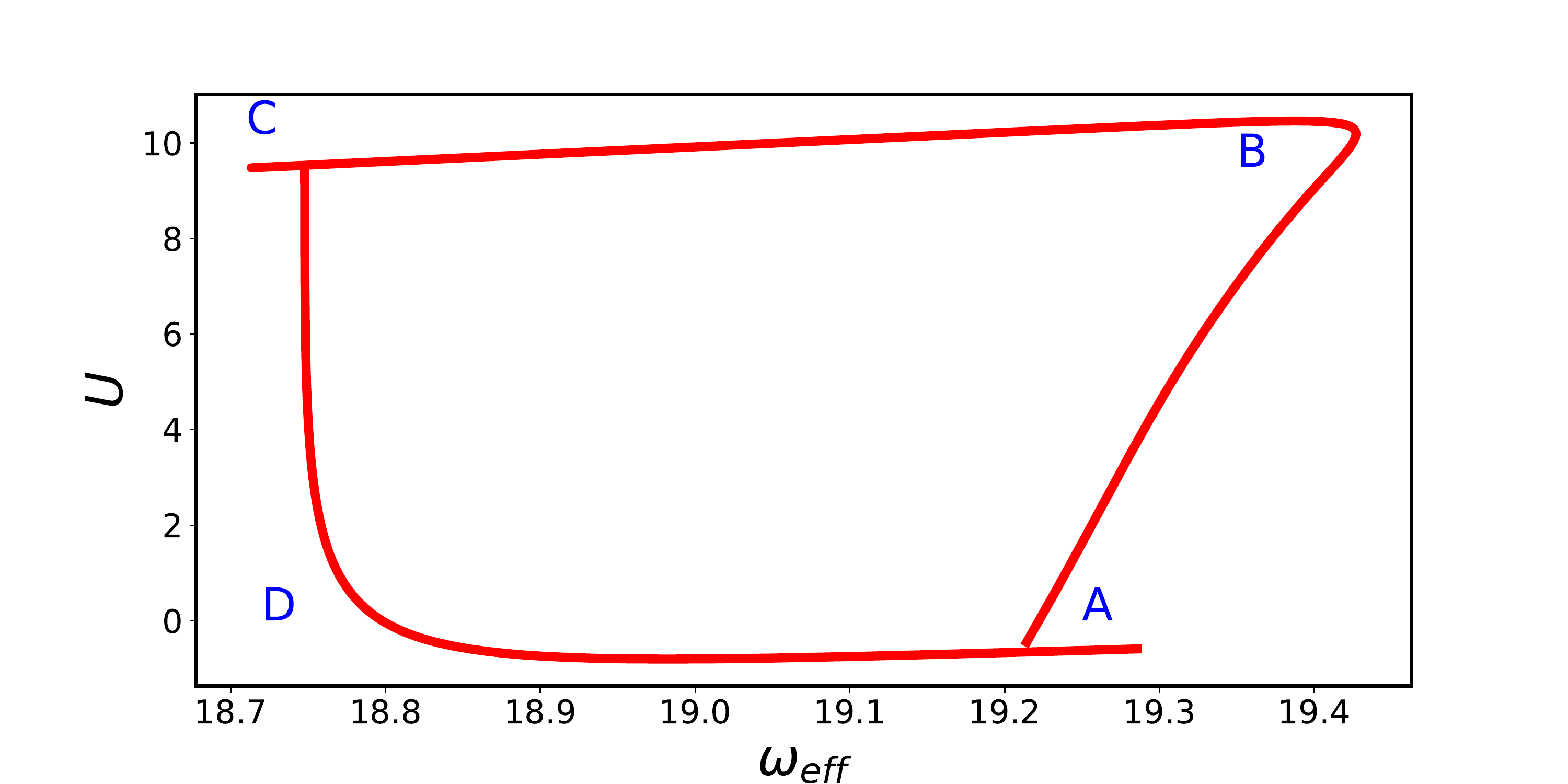}
    \caption{Effective energy-frequency diagram of the radial mode $a$ by considering the dependence of the mean effective energy $U$ to the effective frequency $\omega_{\text{eff}}$ (in units of $\hbar \omega_z$).}
    \label{fig:fig <n> A0}
\end{figure}

For a single ion in an asymmetrical Paul trap~\cite{singleAtom}, the hot bath is simulated by the external white electric-field noise. The cold bath interaction is realized by exposing the ion to the laser cooling beam. According to~\cite{singleAtom,singleAtom02}, the cooling laser is kept on during the process. Despite the continuous coupling to the heat baths, we can identify an Otto engine cycle in our model. We consider the mode $a$ as the working substance and treat the mode $b$ as the engine's flywheel where the work is stored. We replace the flywheel displacement $q_z=\langle \hat b + \hat b^{\dagger}\rangle$ as a number in model $A_0$. The reduced semi-quantum model describes a squeezed oscillator Hamiltonian for mode $a$, whose diagonalization by the Bogoluibov transformation (see~\ref{sec:appA}.) yields an effective frequency of $\omega_{\text{eff}}$, and the effective mean energy can be defined as $U = \langle H_{\text{cm}}^{'}\rangle$. \\ The Otto engine is identified in~Fig.~\ref{fig:fig <n> A0} with four steps~\cite{singleAtom02}.
\par (A$\rightarrow$B) Hot isochore: the effective trapping frequency ($\omega_{\text{eff}}$) is kept constant at $\omega_h$ while contacting with hot bath at temperature $T_h$. The thermalization will be done after a time of $\tau_h$.
\par (B$\rightarrow$C) Adiabatic expansion: The atom evolves in the trapping potential while isolated from the hot bath. After a time $\tau_z$, the effective frequency will be changed from $\omega_h$ to $\omega_c$, and work will be done by considering the displacement in the axial direction. \par (C$\rightarrow$D) Cold isochore: the effective frequency remains in $\omega_c$, and the cold bath with temperature $T_c$ is contacted with the system for a time $\tau_c$. 
\par (D$\rightarrow$A) Adiabatic compression: the effective frequency returns to the initial frequency another time $\tau_z$ in the last step.\\ 
The time scales can be considered by $\tau_h \approx \tau_c < \tau_z$. An effective, finite-time (non-ideal) Otto cycle (ABCD) is defined in the diagram (Fig.~\ref{fig:fig <n> A0}) where we plot dependence of $U$ on $\omega_{\text{eff}}$ in the steady state at $\Bar{n}=3$. 

\subsection{Performance of the engine} \label{sec: performance}
We have seen that the engine operates by converting the thermal energy harvested by the radial mode into coherent oscillations of the axial vibrational mode. We can envision the vibrational mode as the work reservoir storing coherent energy and examine the engine cycle for the radial mode. For that aim, let's consider a semi-classical model of the vibrational interaction by replacing the coherent axial mode operators, in particular $q_z$ operator in the vibrational coupling with a c-number, which is its expectation value. After that, we can apply the Bogoluibov transformation (as described in~\ref{sec:appA}) to the
effective semi-classical model, which yields a simple uncoupled quantum oscillator model. Accordingly, we conclude that the vibrational modes, particularly the radial mode, are Gaussian states under the semi-classical model. More specifically, the radial mode is a Gibbsian (thermal equilibrium) Gaussian state for which a simple Von-Neumann entropy expression can be derived in the form~\cite{SemiQHE} 
\begin{equation}\label{eq:entropy a}
S = (1+\langle\hat n_c\rangle)\ln (1+\langle\hat n_c\rangle) - 
\langle\hat n_c\rangle \ln(\langle\hat n_c\rangle).    
\end{equation}
The effective temperature of the radial mode Gibbsian state is given by
\begin{equation}\label{effective tem}
T_{\text{eff}} = \omega_{\text{eff}}/ \ln(1 + 1/\langle\hat n_c\rangle).     
\end{equation}
Here, the $\langle \hat n_c \rangle$ comes from the diagonalized form of radial mode in Hamiltonian employing Bogoliubov transformation (\ref{sec:appA}).

The temperature-entropy $T-S$ diagram is plotted in Fig.~\ref{fig entropy-tem} which closely resembles an Otto engine. The potential work output can be obtained by calculating the area of the T-S diagram. We approximate the work from the T-S diagram, and the net work is estimated by $W \approx 0.22~\hbar \omega_z \simeq 7.3 \cross 10^{-30}$ joules. The Power can be calculated by dividing work by the cycle time $T = 20 \cross 2\pi/\omega_z$ as $P \simeq 2.9 \cross 10^{-27}$ Watts. The determined heat intake of mode $a$ is $Q_{\text{in}} \simeq 3 \cross 10^{-28}$ joules, and therefore the efficiency is $\eta \simeq 2.4\%$.  

\begin{figure}
    \centering
    \includegraphics[width=8.6cm]{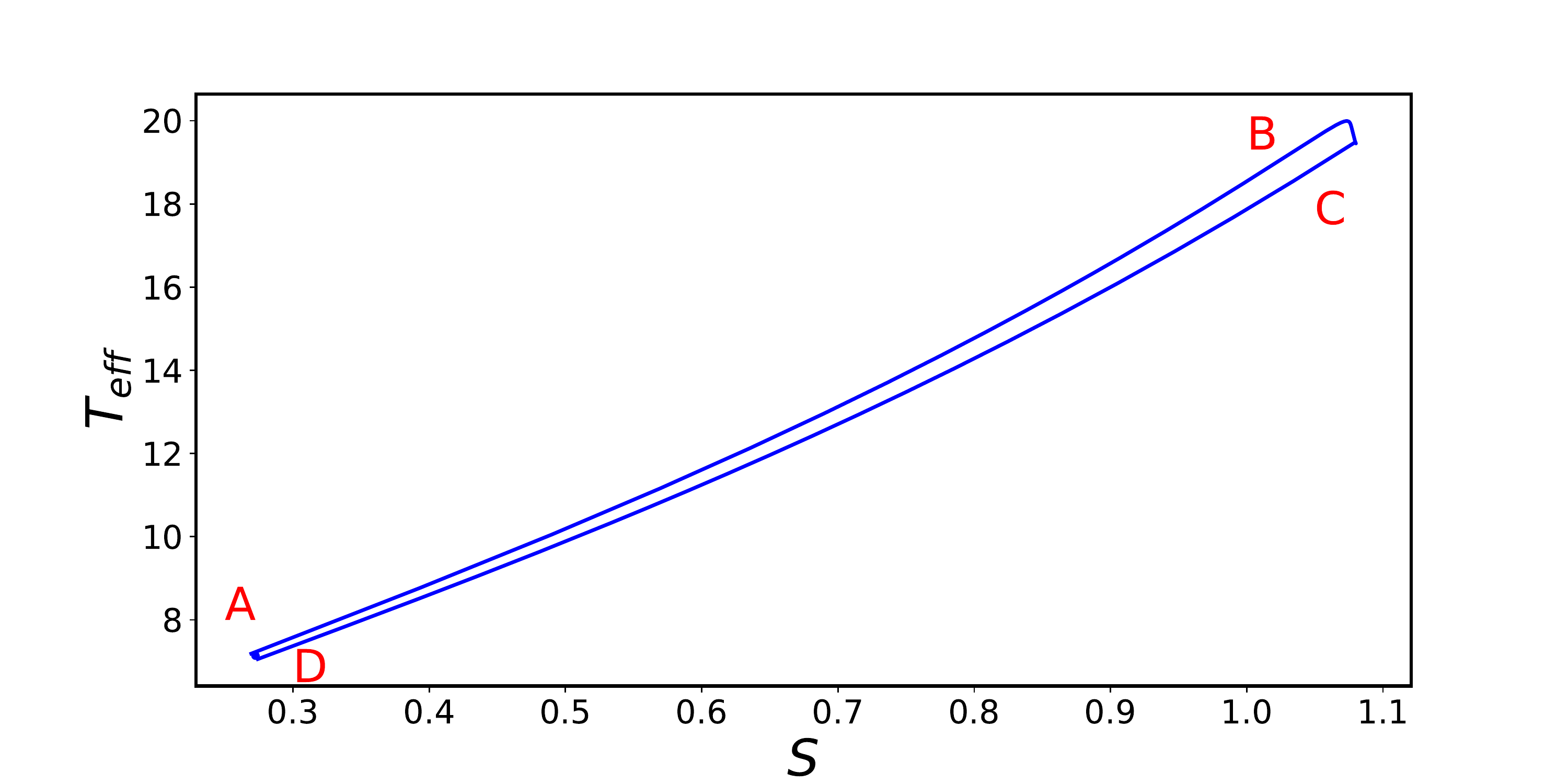}
    \caption{T-S diagram of the mode $a$
which closely resembles that of an Otto engine (in units of $\hbar \omega_z$)}.
    \label{fig:fig entropy-tem}
\end{figure}

\begin{figure}
    \centering
    \includegraphics[width=8.6cm]{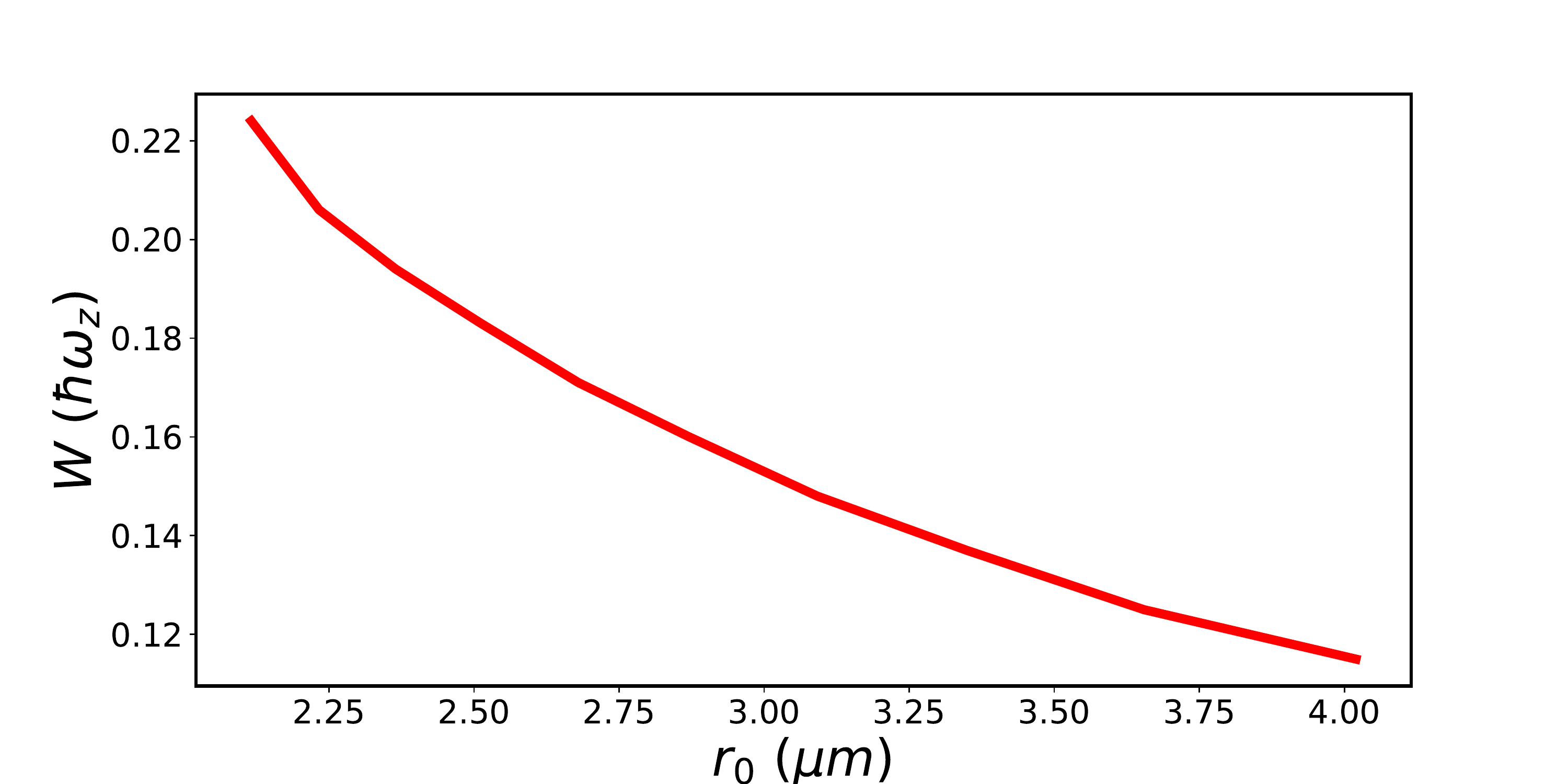}
    \caption{Extracted work $W$ (in unit of $\hbar \omega_z$) versus trap's radius $r_0$ ($\mu$m) where radial and axial frequencies are constant ($\omega_r/2\pi = 1$ MHz, and $\omega_z/2\pi = 50$ kHz) and $\theta = \pi/4$.}
    \label{fig:fig work}
\end{figure}

The coupling coefficient $\beta$ in Hamiltonian~(\ref{eq:ham_CM}) plays an essential role in producing work and achieving higher efficiency, which can be done by revising the trap's geometry in the experiment~\cite{singleAtom}. Engine performance improves by applying the quantum correlations, which are significant only for much higher coupling regimes and at low temperatures. According to the definition of 
$\beta \propto 1/r_0$, we can increase $\beta$ by lowering the trap's radius $r_0$. As we argued in 
Sec.~\ref{sec:parameters}, sufficiently large $\beta$ can be achieved by reducing 
the trap's radius to the micrometer scale.
Fig.~\ref{fig:fig work} shows the dependence of the extracted work, determined by the area of the $T-S$ diagram, on $r_0$. Here, $\theta$, radial, and axial frequencies are kept constant so that $\beta$ increases with $r_0$. It can be pointed out that the work output increases strongly in tightly confined tapered traps. We remark, however, that increasing the coupling coefficient $\beta$ is limited in our treatment where we use a local optomechanical master equation, which requires $\beta$ to be smaller than the radial frequency~\cite{TahirMaster}. 
  
\section{Classical vs Quantum Heat Engine Regimes}\label{sec: classical}

The previous section identifies an effective Otto cycle by treating the radial mode as a 
working fluid and assuming the coherent displacements of axial mode are classical. 
This treatment allows us to use conventional thermodynamic cycle diagrams in terms of mean effective energy, effective temperature, effective frequency (spatial-like parameter), and entropy for our reduced quantum oscillator model for the radial mode. In this case, the potential work output of the effective Otto cycle is computed from the 
$T-S$ diagram. The missing piece of this engine picture is the flywheel, which is the axial mode. The work is translated to the piston (axial) mode, and it is determined in terms of the dissipated power through the axial 
mode, which we will calculate in the present section. Dissipated power is essentially determined with the population of the axial mode (mean number of axial phonons). As we have seen in Sec.~\ref{sec:results}, 
the dynamical equation of the mean number of axial phonons includes a quantum correlator term, which 
can be significant on top of classical correlations when quantum squeezing is present and can enhance the
dissipated power. When there are only classical correlations, we solve the Langevin equations to determine the mean number of axial phonons. In addition, we determine the mean number of axial phonons from the quantum master equation as well. Using these two methods, we will determine the mean number of phonons to be substituted to the same dissipated power formula to compare the classical and quantum dissipated power outputs with each other.

To compare the quantum engine with its classical counterpart, we follow the standard method of neglecting the non-commutative character of the quantum operators (for example, position and momentum operators). In the optomechanical engine models, they are replaced with c-numbers, $\langle \hat{a} \rangle \rightarrow \upsilon_r$ and $\langle \hat{b} \rangle \rightarrow \upsilon_z$~\cite{SemiQHE,power}. The Heisenberg equations of motion are then changed to classical Langevin equations, which further include the stochastic noise terms describing the classical models of the heat baths in the form
\begin{align}
   \Dot{\upsilon}_r &= -i\left(\omega_r \upsilon_r - \frac{\beta}{2} (\upsilon_r^{*} + \upsilon_r) (\upsilon_z + \upsilon_z^{*})\right) \nonumber \\ &- \frac{\kappa_r + \kappa_h}{2} \upsilon_r + \xi_h(t) + \xi_r (t), \nonumber \\ 
    \Dot{\upsilon}_z &= -i\left(\omega_z \upsilon_z - \frac{\beta \upsilon_z}{4} (\upsilon_r^2 + \upsilon_r^{* 2} + 2\upsilon_r \upsilon_r^{*} + 1)\right) \nonumber \\ & - \frac{\kappa_b}{2} \upsilon_z + \xi_z (t).
\end{align}
Here $\xi_k (t)$ presents time dependent delta-correlated and temperature dependent stochastic noise terms where $k = r, z, h$~\cite{SemiQHE,power}.  

Defining the field quadratures $X_i$, $Y_i$ as $\upsilon_i = 1/\sqrt{2}(X_i + i Y_i)$, and writing the noise parameters as $\xi_i = 1/\sqrt{2} (\xi_x^i + i \xi_y^i)$ where $i = r,z$ we can simulate the equations as:
\small
\begin{align}\label{eq: classical engine}
    dX_r & = \left(\omega_r Y_r - \frac{\kappa_a + \kappa_h}{2} X_r\right)dt  + dW_x^h + dW_x^a, \nonumber \\
    dY_r & =  \left( -\omega_r X_r + \sqrt{2}\beta X_r X_z - \frac{\kappa_a + \kappa_h}{2} Y_r \right)dt + dW_y^h + dW_y^a, \nonumber \\
    dX_z &=  \left( \omega_z Y_z - \frac{\kappa_b}{2} X_z \right)dt + dW_x^b, \nonumber \\
    dY_z &=  \left( -\omega_z X_z + \frac{\sqrt{2} \beta}{4} (4 X_r^2 + 1) - \frac{1}{2} \kappa_b Y_z\right)dt + dW_y^b.
\end{align}
\normalsize
Here, $dW^i = 1/\sqrt{2} (dW_x^i + i dW_y^i)$ where $\xi_k^i dt = dW_k^i$, with $k = x,y$, is the Wiener process with width $\sqrt{\kappa_i \Bar{n}_i dt}$. 

\begin{figure}
    \centering
    \includegraphics[width=8.6cm]{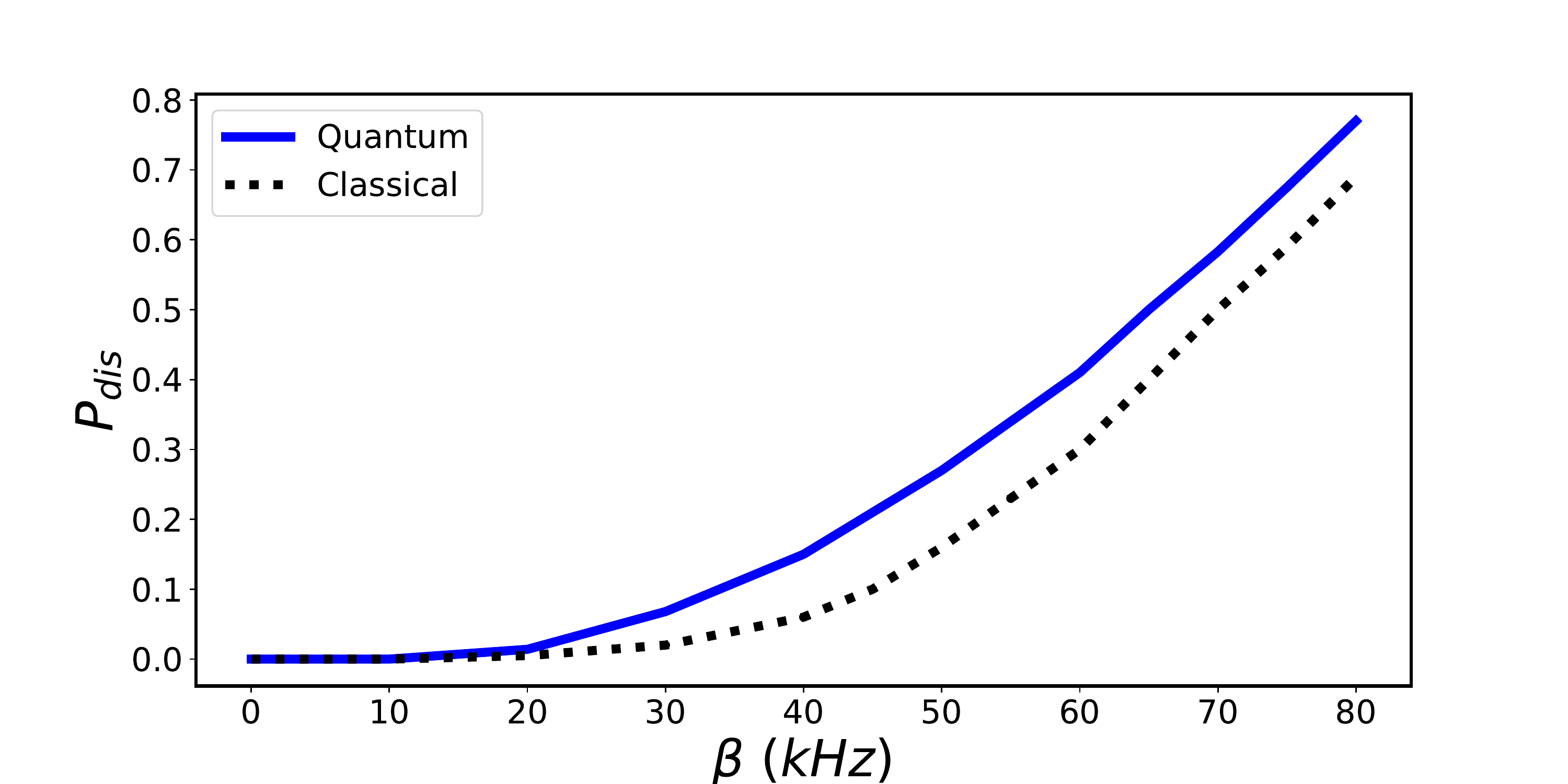}
    \caption{The maximum dissipated power (in units of $\hbar \omega_z $) concerning coupling constant $\beta$ (in units of kHz). The black dashed and solid blue curves show the results of classical and quantum simulations, respectively.}
    \label{fig: classic engine}
\end{figure}

We solve these equations to determine the mean number of phonons,
\begin{align}\label{nz classic}
    \langle \hat{n}_z \rangle = \frac{1}{2}(X_z^2 + Y_z^2),
\end{align}
which is then used to determine the dissipated power. The calculation is then repeated using the quantum master equation instead of the Langevin equations. Fig. \ref{fig: classic engine} shows the maximum dissipated power output of classical and quantum calculations versus $\beta$. It reveals that the quantum model enables power outputs that greatly exceed the power of stochastic engines for large $\beta$, which identifies the regime in which the system can be classified as a genuine quantum heat engine (according to the criteria in Ref.~\cite{quantumSignature}). 

The quantum correlation contributions to the engine performance are negligibly small in the single-ion heat engine experiment~\cite{singleAtom}, and it should be classified as a classical heat engine even at low temperatures according to the criteria of Ref.~\cite{quantumSignature}.
It is necessary to modify the setup by increasing the asymmetry of the tapered trap and decreasing the trap's radius (to increase $\beta$) to bring the single-ion engine of Ref.~\cite{singleAtom} to the domain of quantum heat engines. According to our calculations, the dissipated power output of the engine overcomes the stochastic classical power output with the help of quantum correlations when $\beta \geq 20$ kHz. As an example, to reach this condition by using the parameters defined in Sec.~\ref{sec:results} we can use $\theta = 45^o$ and $r_0 \leq 10$ $\mu$m.

\section{Conclusion}\label{sec:conclusion}
In summary, we examined the single-atom heat engine using the asymmetric trap configuration of experiment Ref.~\cite{singleAtom}~employing a quadratic optomechanical model. In our model, the working fluid mode (radial mode) was simultaneously driven by the contribution of optomechanical, classical, and squeezed terms. We clarified the effects of the coherent and squeezing couplings on the engine's operation beyond the standard optomechanical coupling. Furthermore, it was found that the coupling coefficient $\beta$ plays a critical role in increasing quantum correlations contribution and enables power outputs that greatly exceed the power of stochastic engines.
We showed that the engine should perform in a high coupling and low-temperature regime to benefit from the quantum correlations. According to our calculations, the quantum character of the second order moments of the displacement operator emerges by increasing $\beta$. In the strong coupling regime, the radial mode is in a thermal squeezed state, and as a result, the system treats quantum mechanically. The coupling coefficient depends on the trap's geometry and the frequencies asymmetry, so adjusting the trap's geometry asymmetry and reducing the trap's radius can increase the efficiency and work of the engine. We concluded that a single atom heat engine in a more confined and asymmetric trap operates in the quantum regime where the dissipated power is more significant than its classical counterpart with only classical stochastic correlation.
 
\begin{backmatter}
\bmsection{Acknowledgments}
This work was supported by the Scientific and Technological Research Council of Turkey (TUBITAK) 
with project number 119F200.
\end{backmatter}
\appendix

\section{Effective entropy and temperature of the working mode $a$}\label{sec:appA}

To diagonalize the Hamiltonian (\ref{eq:ham_CM}), we first replace the $q_z$ operator with its expectation value.
The new effective semi-quantum model describes the radial mode as the working fluid of the engine. The effective
reduced model is a quadratic form and it can be diagonalized using the Bogoliubov transformations
\begin{eqnarray} \label{eq:App 01 bogoliubov}
\hat a &=& \hat c \cosh{x} + \hat c^{\dagger} \sinh{x}, \\
\hat a^{\dagger} &=& \hat c \sinh{x} + \hat c^{\dagger} \cosh{x}.
\end{eqnarray}
Here, $\tanh{2x} = \beta q_z / ( 2\omega_r-\beta q_z)$. The diagonalized Hamiltonian becomes
\begin{align} \label{eq:App 02 bogoliubov}
\hat H_{cm}^{'} &= \omega_{\text{eff}} \hat n_c + \omega_z \hat n_z + \omega_r \sinh^2{x} \nonumber \\ & - \frac{\beta q_z }{2} \big (\sinh{2x} + \cosh{2x} \big ), 
\end{align}
where an effective frequency $\omega_{\text{eff}}$ is defined to be
\begin{align} \label{eq:App 03 bogoliubov}
\omega_{\text{eff}} = (\omega_r - \frac{\beta q_z}{2})\cosh{2x} - \frac{\beta q_z }{2} \sinh{2x}.
\end{align}



\bibliography{sample}






\end{document}